\documentclass[preprint,superscriptaddress,preprintnumbers,amsmath,amssymb]{revtex4}
\usepackage{epsfig,graphicx}

\makeatletter
\def\@dotsep{4.5}
\makeatother

\begin{document}

\title{Collective Flow of Protons and Negative Pions in Nucleus-Nucleus Collisions at
Momentum of $4.2 \div 4.5$ AGeV/c}

\author{L.\ Chkhaidze}
\email{ichkhaidze@yahoo.com} \email{ida@hepi.edu.ge}
\affiliation{Institute of High Energy Physics and Informatization, Tbilisi State University, Tbilisi}
\author{P.\ Danielewicz}
\email{danielewicz@nscl.msu.edu}
\affiliation{National Superconducting Cyclotron Laboratory, Michigan State University, East Lansing, Michigan, USA}
\author{T.\ Djobava}
\email{Tamar.Djobava@cern.ch} \email{djobava@hepi.edu.ge}
\affiliation{Institute of High Energy Physics and Informatization, Tbilisi State University, Tbilisi}
\author{L.\ Kharkhelauri}
\affiliation{Institute of High Energy Physics and Informatization, Tbilisi State University, Tbilisi}
\author{E.\ Kladnitskaya}
\affiliation{Joint Institute for Nuclear Research, Dubna, Russia}

\begin{abstract}
Collective flow of protons and negative pions has been studied within the momentum region of
$4.2 \div 4.5$\,AGeV/c ($E =3.4 \div 3.7$\,AGeV) for different projectile-target combinations involving carbon and,
specifically, He-C, C-C, C-Ne, C-Cu and C-Ta.  The data stem from the SKM-200-GIBS streamer chamber
and from Propane Bubble Chamber systems utilized at JINR.  The directed flow of protons grows dramatically
in the carbon region when the counterpart nucleus grows in mass between He and Ta.
The elliptic proton flow points out of the reaction plane and also strengthens as system mass
increases.  Within the reaction plane, the
negative pions flow in the same direction as protons for the lighter of the investigated systems,
He-C, C-C and C-Ne, and in the opposite direction for the heavier, C-Cu and C-Ta.
The Quark-Gluon String Model reproduces observed changes in the flow with system mass.
\end{abstract}

\maketitle

\section{Introduction}

One of the central goals of the high-energy heavy-ion collision research has been the determination
of properties of nuclear matter at densities high compared to that in the ground-state nuclei and
at temperatures high compared to energies per nucleon in the ground-state nuclei.
Among efforts towards that goal, important discoveries made have been those of collective flow effects
in the collisions.  A collective flow is
a motion characterized by space-momentum correlations of dynamic origin.  Different aspects of flow
have been investigated both experimentally and theoretically.  The flow produces asymmetries
associated with the reaction plane, in the particle emission patterns.
Theoretically, those asymmetries can be linked to the fundamental properties of nuclear matter and,
in particular, to the equation of state~(EOS)~\cite{stok1}.  Two types of asymmetries have been
identified.  One has been the directed flow in the reaction plane,
associated with the matter "bouncing-off" within the hot participant
region of overlap between colliding nuclei.  The other has been the squeeze-out of the hot
matter moving perpendicular to the reaction plane from within
the participant region.  With energy increasing into ultrarelativistic values the squeeze-out turns
into an in-plane elliptic flow.

By now, the collective flow effects have been investigated over a wide range of energies, from
tens of MeV/nucleon to 200 GeV/nucleon in the center of mass.  For most part the experiments have
relied on electronic techniques.  A streamer chamber served as a detector in the early Berkeley
experiments and later in Dubna experiments.

Regarding investigative strategy, the reaction plane is the plane within which the centers
of initial nuclei lie, separated in transverse direction by the impact parameter~$\vec{b}$.
Spatial asymmetry in the initial state, associated with the reaction plane, gives rise to asymmetries
in the particle emission patterns.  Within the method of analysis of those asymmetries,
proposed by Danielewicz and Odyniec
 \cite{dan1}, those asymmetries are used to estimate the direction of the reaction plane and the asymmetries
 themselves are assessed in relation to the estimated reaction-plane direction.  In addition, for the analysis of asymmetries, a
 Fourier decomposition of azimuthal particle distributions has been employed \cite{vol},
 relative to the reaction plane,
in the context of particle azimuthal correlations.

Chronologically, the collective flow of charged particles has been first observed experimentally
at the Bevalac
by the Plastic Ball~\cite{gust,doss,gutb1}
 and Streamer Chamber~\cite{ren} collaborations.
The flow continued to be explored at Berkeley and at GSI
\cite{ram,her,lie,kug,bril1,chanc} and further at AGS \cite{bar1,bar2,liu,ogi} and
at CERN/SPS \cite{keitz,nashi,aggar, appel1} accelerators.
The first results from Relativistic Heavy Ion Collider~(RHIC) of BNL have been those on
elliptic flow of charged particles at midrapidity of Au-Au collisions at the
energy of $\sqrt{s} = 130 \, \text{A\,GeV}$~\cite{acker}.

At JINR in Dubna, emulsion techniques have been employed to investigate the directed flow
of projectile spectator fragments in the interactions of O, Ne, Si, S, Kr, Au and Pb
with Ag(Br) targets at beam momenta of 0.95, 4.5, 11.6 and 14.5\,A\,GeV/c~\cite{adam}.
Moreover, the flow of protons and $\pi^-$ mesons has been studied at JINR in the central
C-Ne and C-Cu
collisions at beam momentum of 4.5\,A\,GeV/c \cite{chkha1,chkha2}, using the streamer chamber of the SKM-200-GIBS
collaboration, and in the semicentral C-C \cite{sim1,chkha3} and C-Ta
\cite{sim2}
collisions at beam momentum of 4.2\,A\,GeV/c, using the 2~m Propane Bubble Chamber.
In those sets of flow investigations, on one hand the method of Danielewicz and Odyniec has been
employed~\cite{chkha1,chkha2,chkha3} and, on the other, the method of Fourier expansion
of azimuthal distributions~\cite{sim1,sim2}.  The review~\cite{chkha4} has summarized some of the
results obtained in the streamer and bubble chambers.

In this paper, we concentrate on the collective behavior of protons and $\pi^{-}$ mesons in collisions
where carbon is either a projectile or target and specifically
He-C, C-C, C-Ne, C-Cu and C-Ta collisions at beam momentum of
$4.2 \div 4.5$\,A\,GeV/c, measured either in the JINR streamer or bubble chambers.  Details of the strategy
of quantifying the collective flow effects differ from what the Tbilisi group had employed before \cite{chkha4}.

\section{
Experimental data}

The data combine results obtained within the SKM-200-GIBS set-up and within the 2 m Propane Bubble Chamber of JINR.

The SKM-200-GIBS setup consists of a 2 m streamer chamber placed in the magnetic field
of 0.8\,T and of a triggering system.  The streamer chamber has been exposed to a
beam of C nuclei accelerated in the JINR synchrophasotron
to the momentum of 4.5\,A\,GeV/c (energy of 3.7\,A\,GeV).
The thickness of Cu solid target, in the form of a thin disc, was
0.2~g/cm$^{2}$.  Neon-gas filling of the chamber also served as a nuclear target.
A central trigger was used to select events with no charged projectile
spectator fragments at $p > 3\,\text{GeV/c}$, within a cone of half angle of either
$\Theta_{\text{ch}} = 2.4^\circ$ or $2.9^\circ$, depending on the run.  The trigger efficiency was
99\% for events with a single charged particle in the cone.
The ratio $\sigma_\text{cent}$/$\sigma_\text{inel}$
 (that characterizes the centrality of selected events) is 9$\pm $1\,\% for C-Ne and
21$\pm $3\,\% for C-Cu.  Details of the acquisition techniques of 723 C-Ne and 663 C-Cu interaction data,
and other experimental procedures such as e.g.\ dealing with biases and corrections
in the analysis have been presented in previous publications \cite{ani1,ani2}.
Investigation of
collective flow effects generally requires analyzing collisions event-by-event or exclusively.
In this context, it has been important to put an effort into the
identification of $\pi^{+}$ mesons, the admixture of which
amongst positive charged particles has been about $25\div 27$\,\%.
The identification was carried out on the statistical basis
using two-dimensional ($p^\perp$, $p^{L}$) particle distributions~\cite{chkha5}. It had been
assumed, that $\pi^{-}$ and $\pi^{+}$ mesons contribute to a given cell within the
($p^\perp$, $p^L$) plane with equal probability.  As one detail, the difference in multiplicity
of $\pi^{+}$ and $\pi^{-}$ in each event was required to be less than~3.
After the performed identification, the admixture of $\pi^{+}$ mesons amongst the
protons is estimated to be less than $5 \div 7$\,\%.

 The data on He-C, C-C and C-Ta interactions have been obtained using the 2~meter
Propane Bubble Chamber of JINR.
The bubble chamber was placed in a magnetic field of~1.5~T.
Three Ta-plates $(140 \times 70 \times 1)\,\text{mm}^{3}$ in size and mounted
into the fiducial volume of the chamber at a distance of 93~mm from each
other, served as a nuclear target.  The~procedures for separating out
the He-C and C-C collisions in propane and the processing of the data including particle
identification and corrections have been
described in detail in Ref.~\cite{bond}.  The analysis produced 9737 He-C, 15962 C-C and 2469
unambiguously identified C-Ta
inelastic collision events.  From those,
subsamples of semicentral events have been selected for the flow analysis, by requiring the minimal participant
proton multiplicities of $N_\text{part} \geq 3$, $N_\text{part} \geq 4$ and $N_\text{part} \geq 6$ ,
respectively, for the He-C, C-C and C-Ta collisions.  The target fragmentation products have been identified as
those with momentum $p < 0.3\,\text{GeV/c}$.  The~projectile fragmentation products have been defined
as those characterized by the momentum $p > 3\,\text{GeV/c}$ and angle $\theta < 4^\circ$.
In consequence, 6400 He-C, 9500
C-C and  1620 C-Ta semicentral collision events have separated out from the identified inelastic He-C, C-C and
C-Ta collisions.  The overall cross section ratios $\sigma_\text{cent}$/$\sigma_\text{inel}$ is estimated
to be within the range of $15 \div 50$\,\% for the He-C, C-C and C-Ta collisions.

\section{
Directed flow of
protons}

In determining the directed transverse flow of protons, we have employed the method of
Danielewicz and Odyniec \cite{dan1}.
Most of the data below 4~A\,GeV in the literature have been, actually, analyzed following that method.
The advantage of that method is that it can be employed even at small event statistics such as typical
for film detectors.  The method relies on summation over transverse momenta of selected particles
in the events.

The reaction plane is spanned by the impact parameter vector $\vec{b}$ and the beam axis.
Within the transverse momentum method \cite{dan1}, the direction of $\vec{b}$ is estimated
event-by-event in terms of the vector constructed from particle transverse momenta $\vec{p}_{i}^{\perp}$:
\begin{equation}\label{eq:Q}
  \vec{Q} = \sum_{i=1}^n \omega_i \vec{p}_{i}^{\perp} \, ,
\end{equation}
where the sum extends over all protons in an event.  The summation weight is
$\omega_i = 1$ for $y_i > y_c + \delta$, $\omega_i = -1$ for $y_i < y_c - \delta$ and
$\omega_i = 0$ for $y_c - \delta < y_i < y_c + \delta$, $y_i$ is particle rapidity and
$y_c$ is system c.m.\ rapidity.  Particles around the c.m.\ rapidity, with weak correlations with
the reaction plane, are not included in the reaction-plane determination.  For the
He-C and C-C interactions, we have used $\delta =0.2$ and
for the C-Ne, C-Cu and C-Ta interactions we have used $\delta =0.4$.

When referring a specific proton $j$ to the reaction plane, we estimate the direction of the latter
from the
vector $\vec{Q}$ with contribution of proton $j$ removed,
\begin{equation}\label{eq:Qj}
\vec{Q}_j = \sum_{i \ne j} \omega_i \vec{p}_{i}^{\perp} \, ,
\end{equation}
to eliminate the correlation of the particle with itself, competing with the dynamic effect we are after.
Projection of the transverse momentum of proton $j$ onto the estimated reaction plane
is
\begin{equation}\label{eq:pj}
p_j^{x \prime} = \frac{\vec{p}_j^{\perp}  \cdot \vec{Q}_j}{|\vec{Q}_j|} \, .
\end{equation}
The average in-plane momentum components $\langle p^{x \prime} (y)  \rangle $ can be obtained by averaging
over events the momenta in different rapidity intervals.

Due to finite number of particles used in constructing the vector $\vec{Q}$ in (\ref{eq:Q}), the estimated reaction
plane fluctuates in the azimuth around the direction of the true reaction plane.  Because of those fluctuations,
the average momenta $\langle p^{x \prime}  \rangle$ calculated with the estimated reaction plane get reduced as compared to those
for the true reaction plane $\langle p^{x}  \rangle$:
\begin{equation}\label{eq:pxp}
\langle p^{x \prime} (y) \rangle = \langle p^{x} (y) \rangle  \, \langle \cos{\Phi} \rangle \, .
\end{equation}
Here, $\Phi$ is the angle between the true and estimated reaction plane.  The overall correction factor $k= 1/
\langle  \cos{\Phi} \rangle $, which needs to be applied to $\langle p^{x \prime} (y) \rangle$ in order
to obtain $\langle p^{x} (y) \rangle$, is subject to uncertainty, especially at low multiplicities.
In this work, we evaluate $ \langle \cos{\Phi} \rangle $ from the ratio \cite{dan1,beav}:
\begin{equation}\label{eq:cosfi}
\langle \cos{\Phi} \rangle = \frac{\langle \omega p^{x \prime} \rangle}{\langle \omega p^{x } \rangle}
= \left\langle \frac{\omega \, \vec{p}_j^{\perp} \cdot \vec{Q}_j}{| \vec{Q}_j |} \right\rangle
\left/ \left(  \frac{\left\langle Q^2 - \sum_{i=1}^n \left( \omega p_i^\perp  \right)^2 \right\rangle }{\langle n^2 - n \rangle}  \right)^{1/2} \right.
\, ,
\end{equation}
where $n$ is proton multiplicity in an event.  The $ \langle \cos{\Phi} \rangle $ values obtained for different systems
are listed in Table \ref{tab:info}.

To test the quality of our reaction-plane determination, we have followed the procedure
outlined in Ref.~\cite{dan1}.  Specifically, we have divided randomly each event into two
sub-events and we have constructed vectors $\vec{Q}_1$ and $\vec{Q}_2$ for those sub-events.
The distribution of reaction-plane directions from the sub-events peaks at zero degrees in relative
azimuthal angle and it exhibits a width of $\sigma \approx 50 \div 55^\circ$ depending on reacting system.
The distribution of reaction planes determined from full events, about the true reaction-plane direction, should be about
half as wide as the relative distribution of directions from the sub-events~\cite{dan1,wils}.
The thus-estimated spreads of the estimated reaction-plane directions about the true plane direction,
$\sigma_{0} \simeq \sigma/2 = 25 \div 28^\circ$ depending on a system, are comparable
to values reported
in \cite{jain} and to the values of $\sigma_{0}\approx 23 \div 25^\circ$ arrived at in Kr- and Au-induced
collisions with Ag(Br) targets \cite{adam}.

\begin{figure}
\includegraphics[width=1\linewidth]{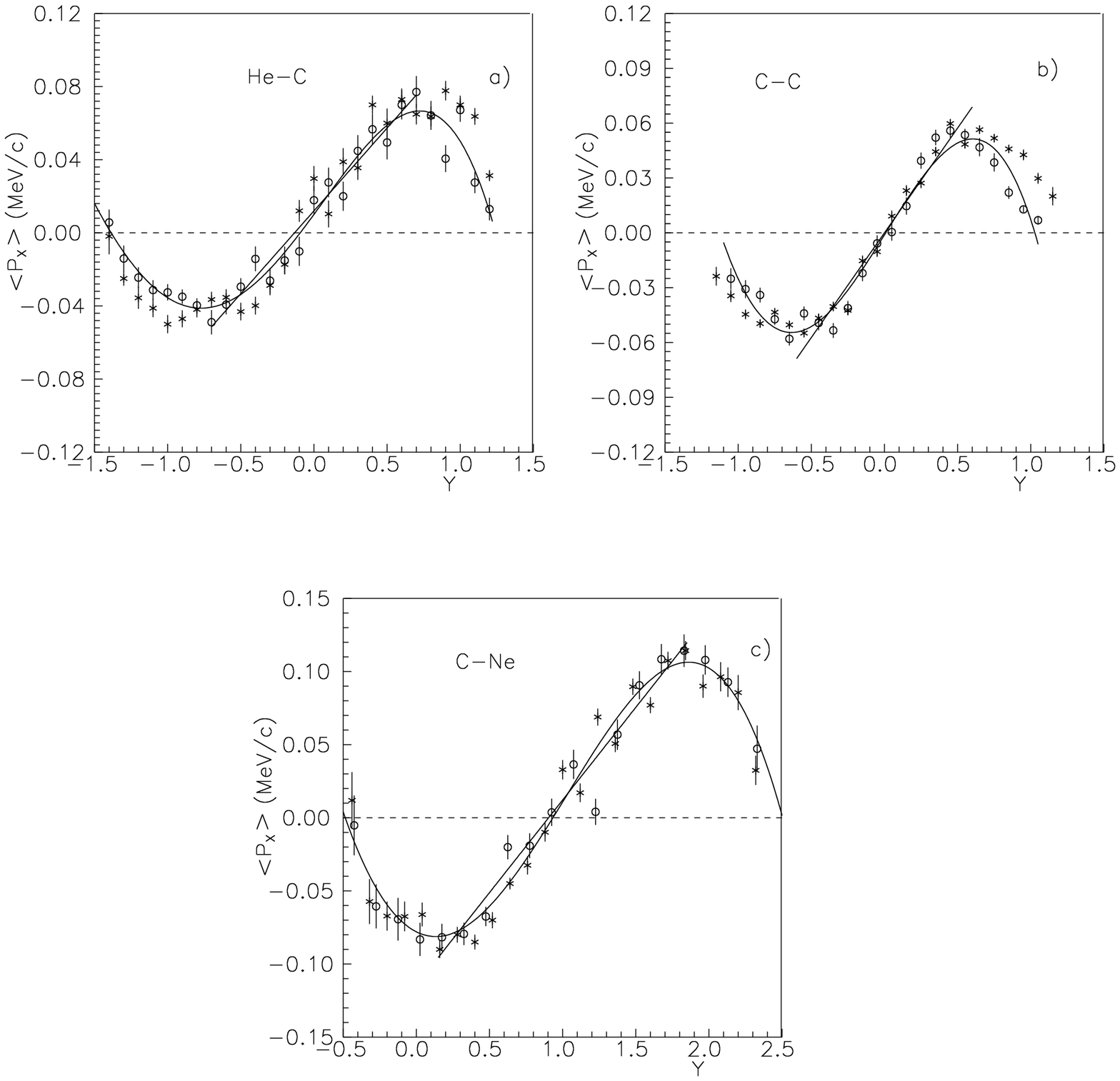}
\caption
{Average component of proton transverse momentum in the reaction plane, as a function of rapidity in the c.m.\ system of
He-C (a) and C-C (b) collisions and in the lab.\ system of C-Ne~(c) collisions.  The data, corrected
for $\langle \cos{\Phi} \rangle$ from Table \ref{tab:info}, are represented by circles.
Results from the QGSM model are represented by crosses.  Straight-line stretches represent
the slope of data at midrapidity cross-over, obtained by fitting the data with a third-order polynomial
within the rapidity region of $-0.70 < y < 0.70$ for He-C, $-0.60 < y < 0.60$ for C-C and $0.15 < y < 1.85$ for~C-Ne.
The curved lines guide the eye over data.}
\label{fig:fig1}
\end{figure}

\begin{figure}
\includegraphics[width=.5\linewidth]{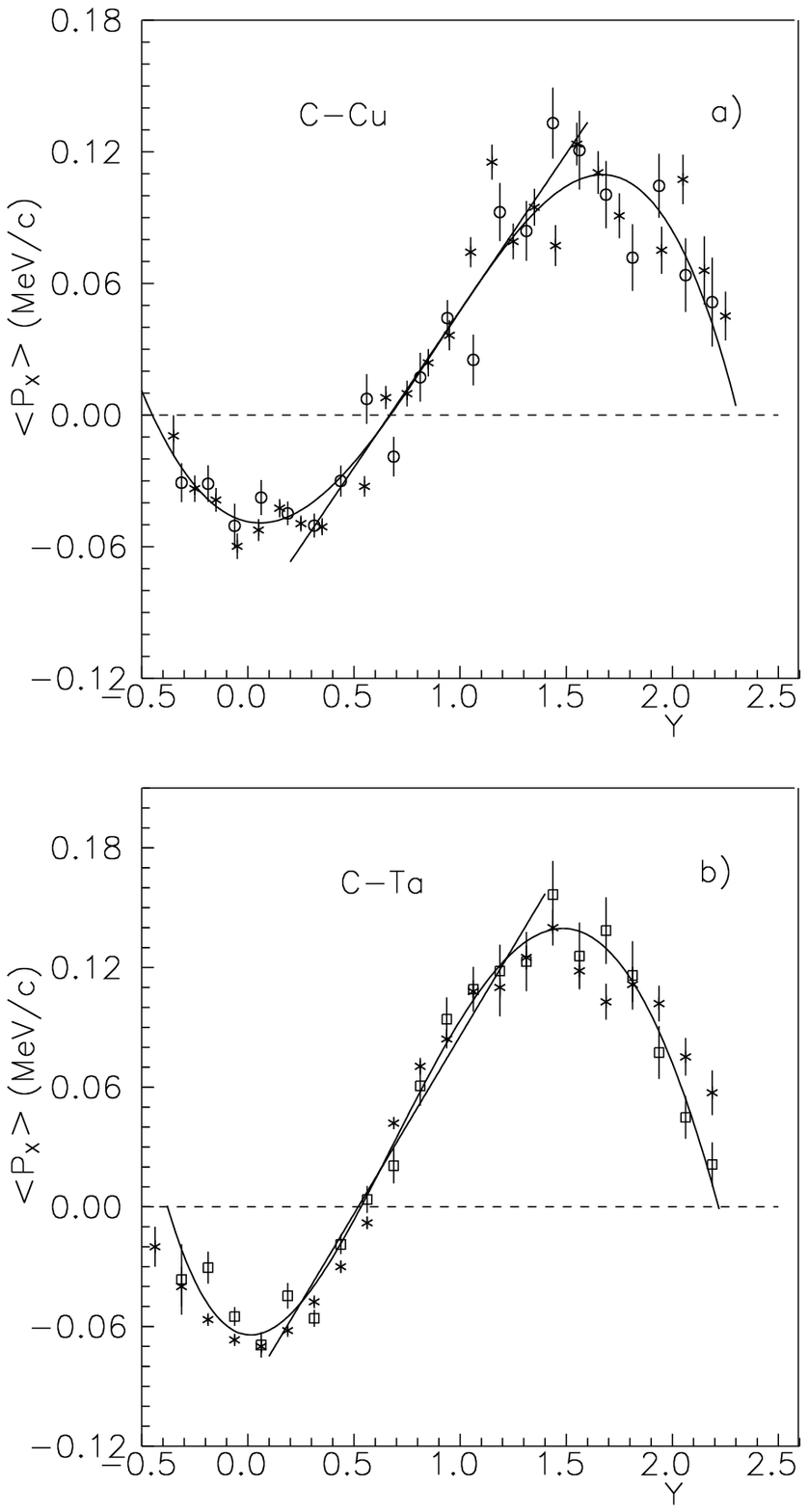}
\caption
{Average component of proton transverse momentum in the reaction plane, as a function of rapidity
in the lab.\ system of C-Cu (a) and C-Ta (b) collisions.  The data, corrected
for $\langle \cos{\Phi} \rangle$ from Table \ref{tab:info}, are represented by circles.
Results from the QGSM model are represented by crosses.  Straight-line stretches represent
the slope of data at midrapidity cross-over, obtained by fitting the data with a third-order polynomial
within the rapidity region of $0.20 < y < 1.60$ for C-Cu and $0.10 < y < 1.40$ for
C-Ta.  The curved lines guide the eye over data.}
\label{fig:fig2}
\end{figure}

With regard to the rapidity, the results of
our analysis of the He-C and C-C interactions are presented in the c.m.\
system and those of the C-Ne, C-Cu and C-Ta interactions are presented in the lab.\ system.
The values of proton $ \langle p^{x} (y) \rangle $ extracted
from the collision events at $4.2 \div 4.5$ A\,GeV/c ($3.4 \div 3.7$ A\,GeV),
corrected for $ \langle \cos{\Phi} \rangle $ from (\ref{eq:cosfi}),
are shown in Fig.~\ref{fig:fig1} for He-C (panel a), C-C (b) and C-Ne (c) and in Fig.~\ref{fig:fig2}
for C-Cu (a) and C-Ta~(b), respectively.  From within this set, the He-C system is the lightest one
within which the directed proton flow has ever been observed.  The maximal values of $ \langle p^{x} (y) \rangle $ in the
carbon region are observed to grow gradually and monotonically as the mass of the counterpart nucleus
increases.  Interestingly, the maximal values of $ \langle p^{x} (y) \rangle $ are always larger in the
region of the lighter of two colliding nuclei.

\begin{figure}
\includegraphics[width=.5\linewidth]{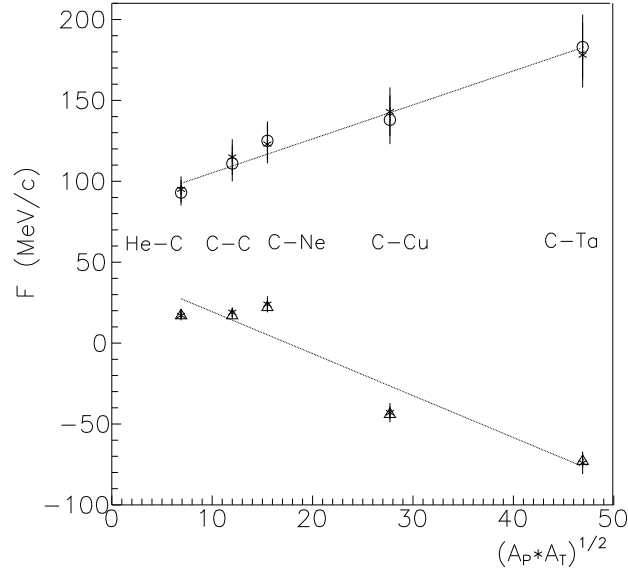}
\caption
{Dependence of the slope $F$ (\ref{eq:Fdef}) on geometric mean of projectile and target masses $(A_P*A_T)^{1/2}$, for the systems indicated in the figure.  The data
are represented by circles and triangles for protons and pions, respectively.  The lines represent linear fits to the data.
Results of the QGSM model are represented by stars in the top and bottom part of the figure, for protons and pions, respectively.
}
\label{fig:fig3}
\end{figure}

One commonly employed measure, quantifying the dependence of $ \langle p^{x} \rangle $ on rapidity, which makes no
distinction between the two colliding nuclei, is the slope~\cite{doss} of $ \langle p^{x} (y) \rangle $ at its midrapidity
cross-over,
\begin{equation}\label{eq:Fdef}
  F = \left. \frac{\text{d}\langle p^{x} \rangle}{\text{d}y} \right\vert_{\langle p^{x} \rangle = 0} \, .
\end{equation}
The slopes, represented in Figs.\ \ref{fig:fig1} and \ref{fig:fig2} by straight lines, have been obtained, as
common, through third-order polynomial fits to midrapidity data, specifically from the rapidity intervals of
$-0.70 \div 0.70$ for He-C, $-0.60 \div 0.60$ for C-C, $0.15 \div 1.85$ for C-Ne, $0.20 \div 1.60$ for C-Cu
and $0.10 \div 1.40$ for C-Ta.  The slopes, corrected for $\langle \cos{\Phi} \rangle $, are further provided
in Table \ref{tab:info} and they are plotted
against the geometric mean of projectile and target masses, in~Fig.~\ref{fig:fig3}.

In Fig.~\ref{fig:fig3} and Table~\ref{tab:info}, it is seen that the proton $F$ rises monotonically with the rise
in the mass of the nucleus counterpart to C, from $F^p = 95 \pm 8$~MeV/c for He-C to $F^p = 178 \pm 20$~MeV/c for C-Ta.
Within the observed mass range and errors, the dependence of`$F^p$ on the system mean geometric mass can be described
as linear.  If a power law forced through zero at zero mass were assumed for the dependence of $F^p$ on mass, though,
a`fractional power would have resulted.  A deviation from linearity is further anticipated as
$F^p$ would need to saturate with mass in the hydrodynamic limit of large masses.

The experimental results on proton directed flow have been compared to the predictions of the
Quark-Gluon String Model (QGSM).
Detailed description of that model can be found in~\cite{amel1,amel2}.
The QGSM is based on the Regge and string
 phenomenology of particle production in inelastic binary hadron collisions.
The model oversimplifies nuclear effects in that nucleon mean-field effects are ignored as well as nucleon
coalescence into clusters.  Within~QGSM, nuclear densities are used for selecting coordinates of original
nucleons.  This is followed by the formation of quark-gluon strings which fragment into hadrons.  Those hadrons
rescatter.  In~QGSM, the sole cause of sidewards flow is the hadron rescattering.  In~all other models in the literature, where
produced pressure comes from rescattering alone, the flow in the specific energy regime was found significantly underestimated, albeit
predominantly in heavier systems.  However, in collisions involving carbon studied here, QGSM turned out to describe the flow rather well.

For simulating the model events, we have employed the COLLI Monte-Carlo generator~\cite{amel3} based on QGSM.
To the generated events, a detector filter has been applied and, in the case of C-Ne and C-Cu collisions, a trigger filter.
In mimicking, in particular, the deterioration of experimental efficiency for registering vertical tracks, protons characterized
by deep angles greater than $60^\circ$ have been excluded from an analysis.
In the past, it has been found that the applied filters selected peaked distributions of impact parameters for the
collision events, characterized by the average $b$-values of:
$\langle b \rangle = 2.80$~fm and $\langle b \rangle = 2.65$~fm for semicentral He-C and C-C events \cite{chkha3},
respectively, $\langle b \rangle = 2.20$~fm and $\langle b \rangle = 2.75$~fm for central C-Ne and C-Cu \cite{chkha1,chkha2},
respectively, and $\langle b \rangle = 6.54$~fm for C-Ta semicentral.  For the present study, all events have been generated
at fixed $b$-values equal to the above averages, with event numbers given in Table \ref{tab:info}.
Subsamples of those events have been chosen for flow analysis
following the same multiplicity criteria as for the data.  The proton flow results from QGSM are superimposed on data in
Figs.~\ref{fig:fig1} and \ref{fig:fig2}.  As can be seen, the model describes the data there rather well.  The mass dependence, as seen in
Fig.~\ref{fig:fig3} is, in particular, fairly well reproduced.

\section{Proton Elliptic Flow}

Another flow identified in the literature, persisting even around midrapidity where the directed flow is suppressed,
has been the elliptic flow.  The elliptic flow is associated with the shape of participant region of overlap between
the two nuclei, elliptic in the directions transverse to beam axis.  Gradients within the elliptic participant region
are stronger in the direction of the reaction
plane than out of the plane.  That anisotropy favors overall expansion of matter in the direction of the plane.
However, at sufficiently low energies, when
cold spectator matter lingers in the vicinity of an expanding participant matter, the participant expansion gets
shadowed in the direction of the reaction plane and the so-called squeeze-out,
emission out of the reaction plane dominates.  Details depend
on beam energy and on the pace of expansion of the participant matter.  The energy region for our data is particularly interesting
because it is in the vicinity of transition from squeeze-out to in-plane elliptic flow.  Given that different particles may be
emerge at different times, different elliptic flow directions can principally result for different particles.

\begin{figure}
\includegraphics[width=1\linewidth]{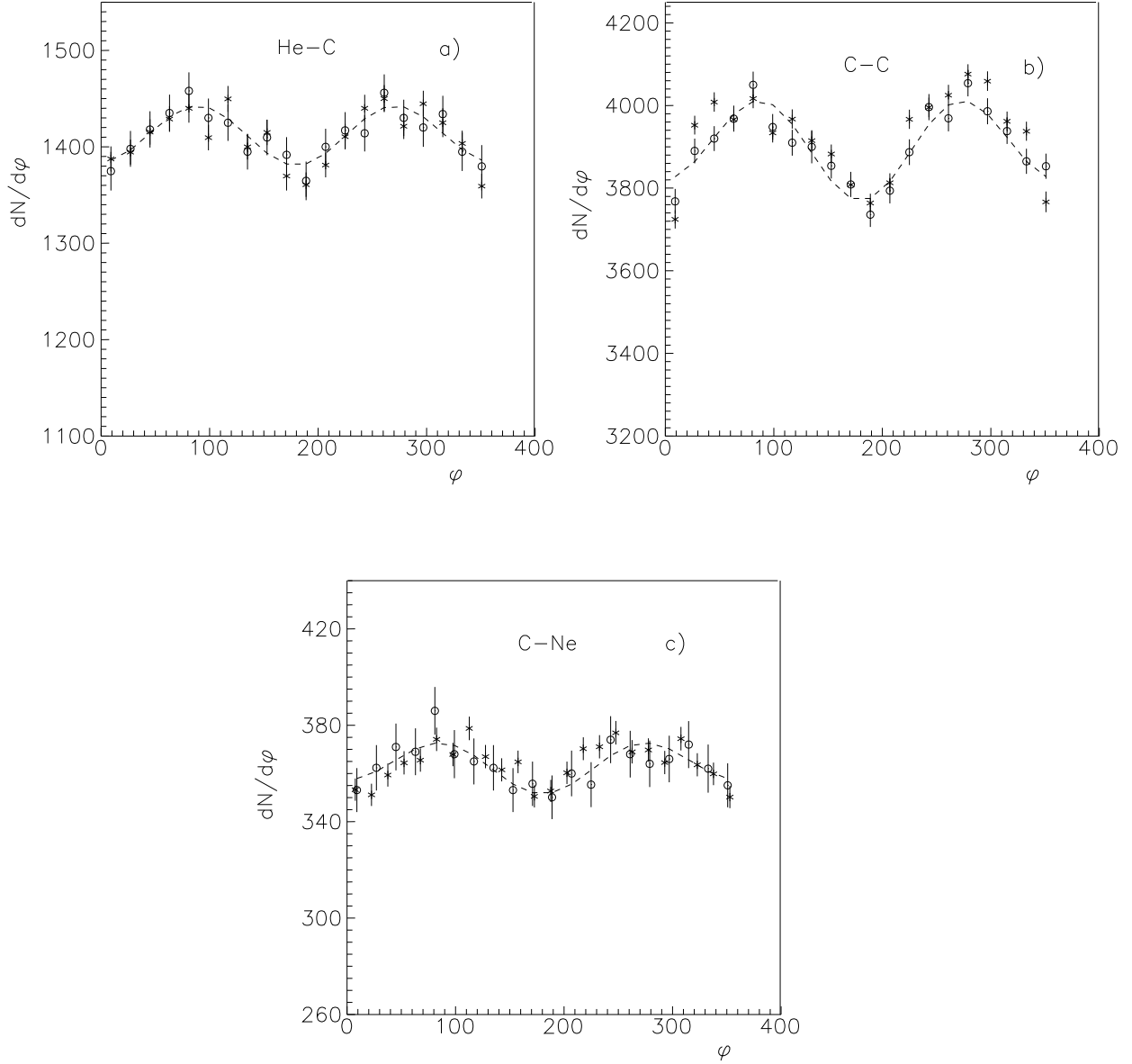}
\caption{Azimuthal distribution of protons relative to the estimated reaction-plane
within the c.m.\ rapidity region of $-1.3 < y < 1.3$ in He-C (a) and C-C (b) collisions and within the lab.\
rapidity region of $-0.4 < y < 2.1$ in C-Ne collisions~(c).
The circles represent data for 20 bins of azimuthal angle.
The lines represent fit to the data with the function
$\text{d} N/ \text{d} \varphi = a_{0}( 1+a_{1}' \cos{\varphi} +a_{2}' \cos2{\varphi})$.
The stars represent distribution calculated within the QGSM model,
with a normalization set to match that of the data.}
\label{fig:fig4}
\end{figure}

\begin{figure}
\includegraphics[width=.5\linewidth]{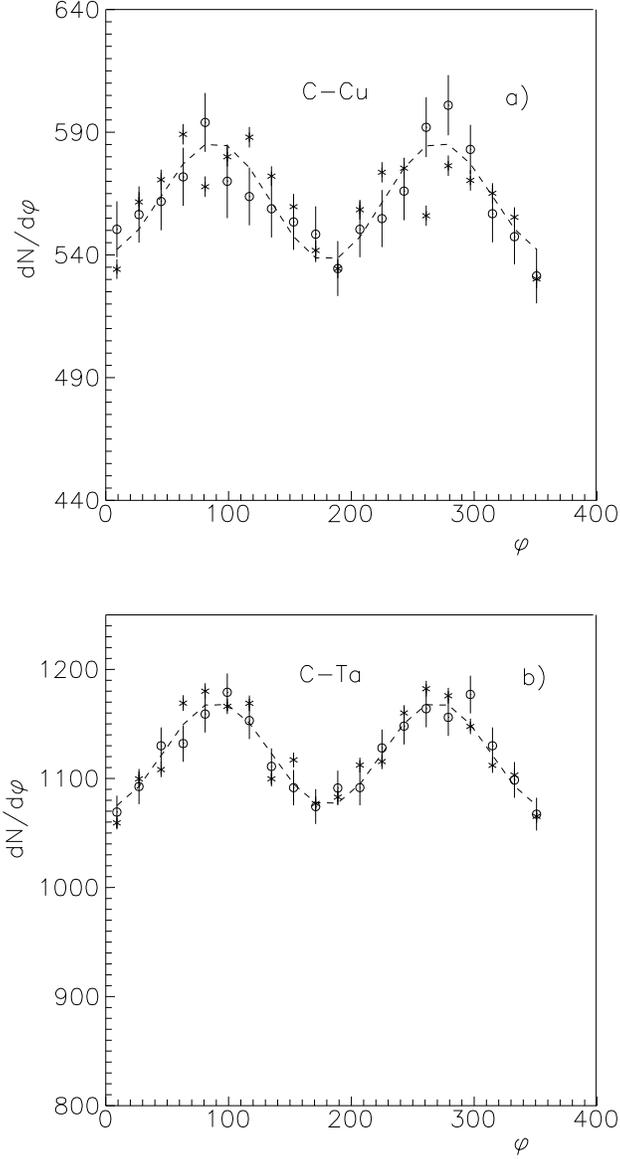}
\caption{Azimuthal distribution of protons relative to the estimated reaction-plane
within the lab.\ rapidity region of $-0.4 < y < 2.1$ in C-Cu (a) and $-0.3 < y < 2.0$ in C-Ta collisions~(b).
The~circles represent data for 20 bins of azimuthal angle.
The lines represent fit to the data with the function
$\text{d} N/ \text{d} \varphi = a_{0}( 1+a_{1}' \cos{\varphi} +a_{2}' \cos2{\varphi})$.
The stars represent distribution calculated within the QGSM model,
with normalization set to match that of the data.}
\label{fig:fig5}
\end{figure}

Azimuthal distributions of protons, relative to the reaction plane direction estimated with (\ref{eq:Qj}), are shown, respectively,
in Fig.~\ref{fig:fig4} for He-C (a), C-C (b) and C-Ne (c) collisions and in Fig.~\ref{fig:fig5} for C-Cu (a) and C-Ta (b) collisions.
The azimuthal angle is one for which $\cos{\varphi_j} = p_j^{x \prime}/ {p}_j^{ \perp}$.
The distributions exhibit a clear modulation, with maxima occurring around 90$^\circ$ and 270$^\circ$, i.e.\ out of the reaction plane,
and minima around 0$^\circ$ and 180$^\circ$, i.e.\ in the reaction plane.  This is the pattern such as observed at lower rather than
higher energies.

Notably, the fluctuations of the true reaction plane about the estimated one tend to flatten the azimuthal distributions of the
particles relative to the plane.
To quantify further the observed modulation, we fit the azimuthal distributions with the Fourier cosine-expansion (given
the system invariance under reflections with respect to the reaction plane):
\begin{equation}\label{eq:fco}
  \frac{\text{d} N}{ \text{d} \varphi} = a_0 \, \left(  1 + a_1' \, \cos{\varphi} + a_2' \, \cos{2 \varphi} \right) \, .
\end{equation}
The squeeze-out signature is the negative value of the coefficient $a_2'$.  Compared to the coefficient $a_2$ associated with a
distribution relative to the true reaction plane, that coefficient is reduced, though, by \cite{pink,andr}
\begin{equation}
\label{eq:a2red}
  a_2' = a_2 \, \, \langle \cos{2 \Phi} \rangle \, ,
\end{equation}
in an analogy to the reduction in (\ref{eq:pxp}).  The reduction coefficient for (\ref{eq:a2red}) may be estimated
from
\begin{equation}
\label{eq:c2f}
  \langle \cos{2 \Phi} \rangle = \frac{| \langle (p^{x \prime })^2 - ( p^{y \prime})^2 \rangle  |}{| \langle (p^{x  })^2 - ( p^{y })^2 \rangle  |} \, ,
\end{equation}
where the numerator and denominator on the r.h.s.\ are, respectively, obtained from
\begin{equation}
  \langle (p^{x \prime })^2 - ( p^{y \prime})^2 \rangle = \bigg\langle 2 \bigg(
  \frac{\vec{p}_j^\perp \cdot \vec{Q}_j}{Q_j} \bigg)^2 - (p_j^\perp)^2 \bigg\rangle
  \, ,
\end{equation}
and
\begin{equation}
| \langle (p^{x })^2 - ( p^{y })^2 \rangle | =
\sqrt{ \frac{ \langle 2 \overline{\overline{T}} : \overline{\overline{T}} - \sum_{i=1}^n (p_i^\perp)^4    \rangle   }{\langle n^2 - n \rangle }   \, .        }
\end{equation}
In the above, the transverse tensor $\overline{\overline{T}}$ is
\begin{equation}
T^{\alpha \beta} = \sum_{i=1}^n \left( p_i^\alpha \,  p_i^\beta -
\frac{1}{2} (p_i^\perp)^2 \, \delta^{\alpha \beta} \right) \, , \hspace*{2em} \alpha=x,y \, ,
\end{equation}
and
\begin{equation}
\overline{\overline{T}} : \overline{\overline{T}} = \sum_{\alpha,\beta=x}^y T^{\alpha \beta} \, T^{\alpha \beta}
= \left( T^{xx} \right)^2 + \left( T^{yy} \right)^2 + 2 \left( T^{xy} \right)^2  \, .
\end{equation}

Use of Eq.~(\ref{eq:c2f}) for a reaction requires that some elliptic anisotropy is present in the analyzed system to start with.
Values of $\langle \cos{2 \Phi} \rangle $ estimated using Eq.~(\ref{eq:c2f}) for our systems fall within the range of $0.520 \div 0.582$,
depending on the system,
and are listed in Table~\ref{tab:info}.
The fits to the azimuthal distributions with Eq.~(\ref{eq:fco}) are illustrated with solid lines in Figs.~\ref{fig:fig4}
and~\ref{fig:fig5}.  The elliptical modulation parameters, corrected according
to Eq.~(\ref{eq:a2red}), from the fits made under different cuts to analyzed particles, are provided in Table~\ref{tab:eli}.
The elliptic anisotropy, quantified in terms of the $a_2$-coefficient, strengthens as the mass of a nucleus counterpart to carbon increases
and as the transverse momentum increases.  The ratio of out-of-plane to in-plane emission probabilities can be estimated as
\begin{equation}\label{eq:R=}
 R=\frac{1-a_{2}}{1+a_{2}} \, ,
\end{equation}
with values also provided in Table \ref{tab:eli}.  Regarding kinematic region, the elliptic anisotropy is seen to
strengthen in Table \ref{tab:eli} as the cut around midrapidity region narrows.

In the past, the elliptic flow has been investigated by different groups in different systems for different particles.
Thus, the Plastic Ball \cite{gutb2,kamp} and FOPI \cite{pelte} groups have investigated the flow in Ca+Ca, Nb+Nb,
Ni-Ni, Xe-CsI, Au-Au collisions from 0.15 to 1.0 GeV/nucleon, for protons,
light fragments and $\pi^{\pm}$ mesons.  The KaoS \cite{bril2} group has investigated the flow in Au+Au and Bi+Bi at 0.4-1.0 GeV/nucleon,
for protons, light fragments, pions and kaons.  Regarding theoretical descriptions, those lower-energy data have been, in particular, successfully described
by the IQMD model \cite{hart}.  Our results, such as of strengthening of the flow with increase in system mass or transverse
momentum, confirm the earlier experimental findings, some from lower and some from higher \cite{pink,andr} energies.

Our results on proton elliptic flow have been further compared to those from the QGSM model of the reactions.
Proton distributions relative to the reaction
plane estimated as in the experiments, normalized as in the experiments, are overlaid over the data in Figs.~\ref{fig:fig4}
and~\ref{fig:fig5}.  The $a_2$-coefficient values from the QGSM model, for distributions relative to the known
reaction plane, are further given in Table \ref{tab:eli}.  As may be seen, the calculations compare favorably to data.
On the other hand, in our testing this model appears to overestimate
the strength of elliptic flow in the heavier Au + Au system, when the higher-energy E-895 data~\cite{pink,andr}
are interpolated across our energy domain.

At higher energies, the elliptic flow results have been mostly expressed in terms of the average cosine \cite{pink,andr,bar3,luk05,appel}
\begin{equation}\label{eq:v2=}
  v_2 = \langle \cos{2 \phi} \rangle \, ,
\end{equation}
where $\phi$ is the angle relative to the true reaction plane.  Given the Fourier expansion (\ref{eq:fco}),
the average $v_2$ differs from the coefficient $a_2$ by just a factor:
\begin{equation}\label{eq:va2}
  v_2 = a_2/2  \, .
\end{equation}

\begin{figure}
\includegraphics[width=.5\linewidth]{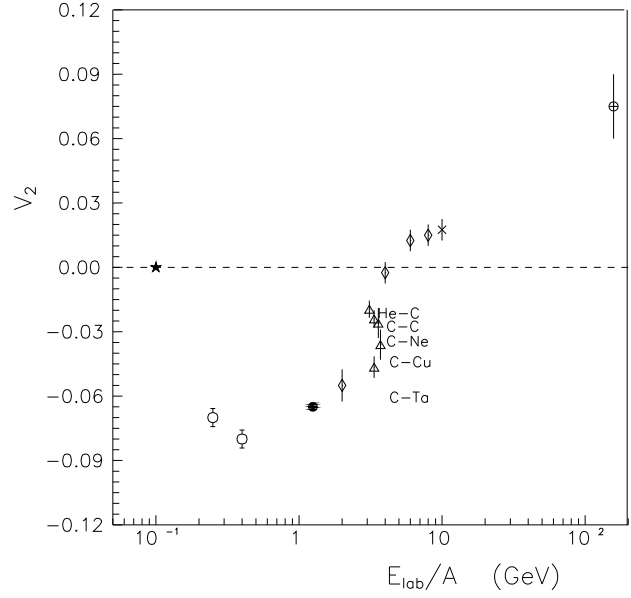}
\caption{Excitation function of elliptic flow in terms of the $v_2$-coefficient vs laboratory energy per nucleon.
The triangles represent our results slightly offset in energy to make it easier to discern the results for different systems.
The symbols representing results of other collaborations, for Au-Au, are:
star - INDRA~\cite{luk05}, open circles - Plastic Ball~\cite{gutb1}, filled circle - EOS ~\cite{pink}, diamonds - E895~\cite{pink},
diagonal cross - E877~\cite{bar3} and circled cross - NA49~\cite{appel}.
}
\label{fig:fig6}
\end{figure}

Most thoroughly and over the widest energy range, the elliptic flow has been studied for protons in Au+Au collisions.
Figure~\ref{fig:fig6} compares our results to the systematics for Au+Au.  The flow generally changes sign just above our bombarding
energy.  It is likely, in the view of our results, though, that the transition energy depends on the colliding system.  It should be mentioned
that in the Au+Au measurements, due in particular to higher particle multiplicity, it has been possible to achieve
a more narrow and more central relative impact parameter selection than for our systems.  Determination of excitation functions for different-size
systems could facilitate separation of the effect of transport properties of nuclear medium and of the equation of state
and facilitate localization of the transition to quark-gluon plasma~\cite{oli}.

\section{Directed Flow of Negative Pions}

Pions, being copious in relativistic heavy-ion collisions, can probe the reaction dynamics independently of nucleons.  Of interest,
in particular, is the relation between the directed flow of nucleons and pions.

Historically, the pattern of pion emission relative to the reaction plane has been first studied at
the Bevalac by the Streamer Chamber group \cite{dan3,keane} and later by the EOS collaboration~\cite{kint} at Saturne by the Diogene group~\cite{gos1}.
The Diogene group investigated, in~particular, collisions similar to ours where one nucleus, in their case the projectile Ne, was fixed
while the counterpart target nuclei were varied.
  In the investigations at Bevalac and Saturne, projection of pion transverse momentum onto the reaction plane has been examined.

\begin{figure}
\includegraphics[width=1\linewidth]{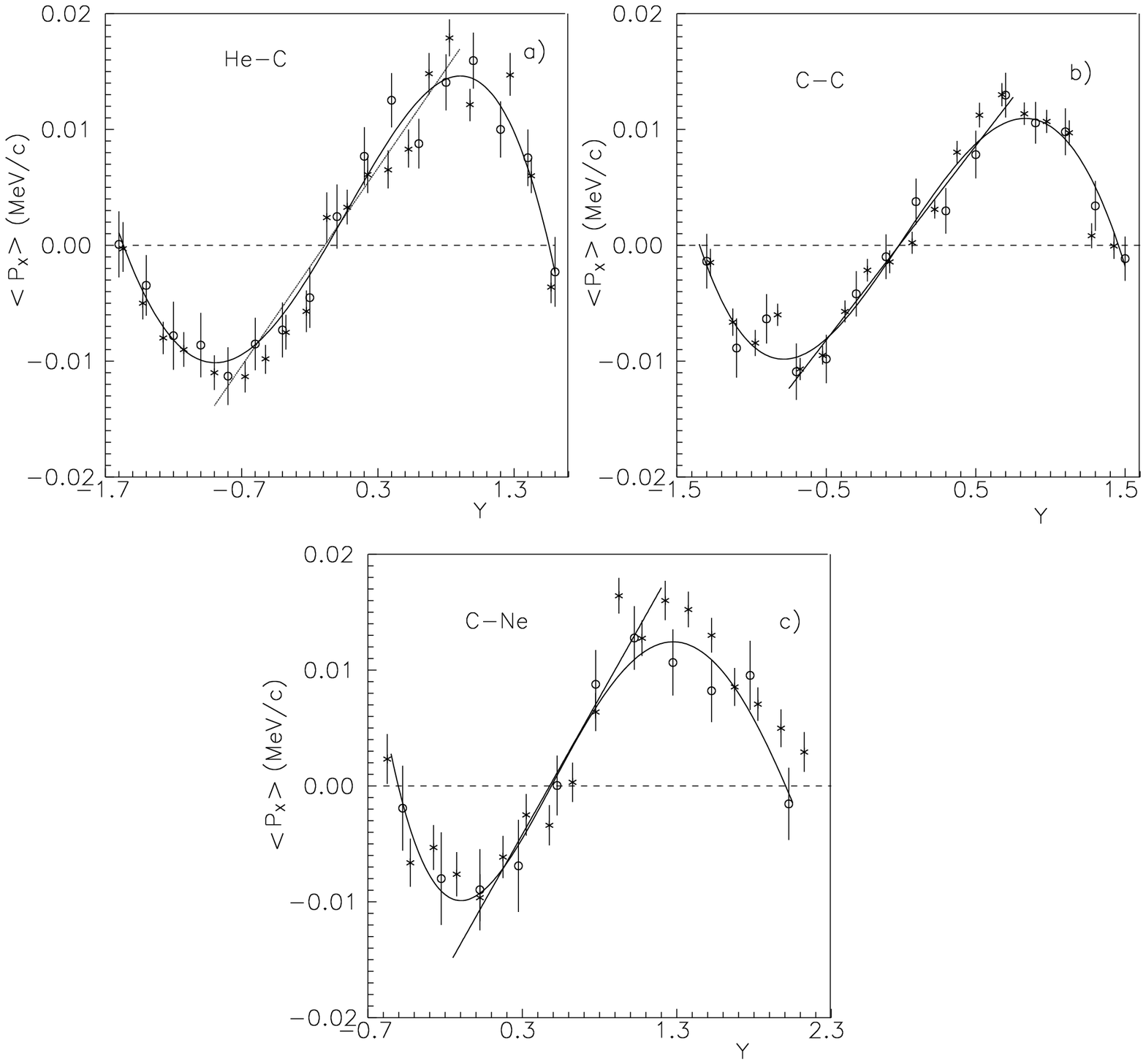}
\caption
{Average component of $\pi^-$-transverse momentum in the reaction plane, as a function of rapidity in the c.m.\ system of
He-C (a) and C-C (b) collisions and in the lab.\ system of C-Ne (c) collisions.  The data, corrected
for $\langle \cos{\Phi} \rangle$ from Table \ref{tab:info}, are represented by circles.
Results from the QGSM model are represented by crosses.  Straight-line stretches represent
the slope of data at midrapidity cross-over, obtained by fitting the data with a third-order polynomial
within the rapidity region of $-0.90 < y < 0.90$ for He-C, $-0.75 < y < 0.75$ for C-C and $0.15 < y < 1.20$ for~C-Ne.
The curved lines guide the eye over data.}
\label{fig:fig7}
\end{figure}

\begin{figure}
\includegraphics[width=.5\linewidth]{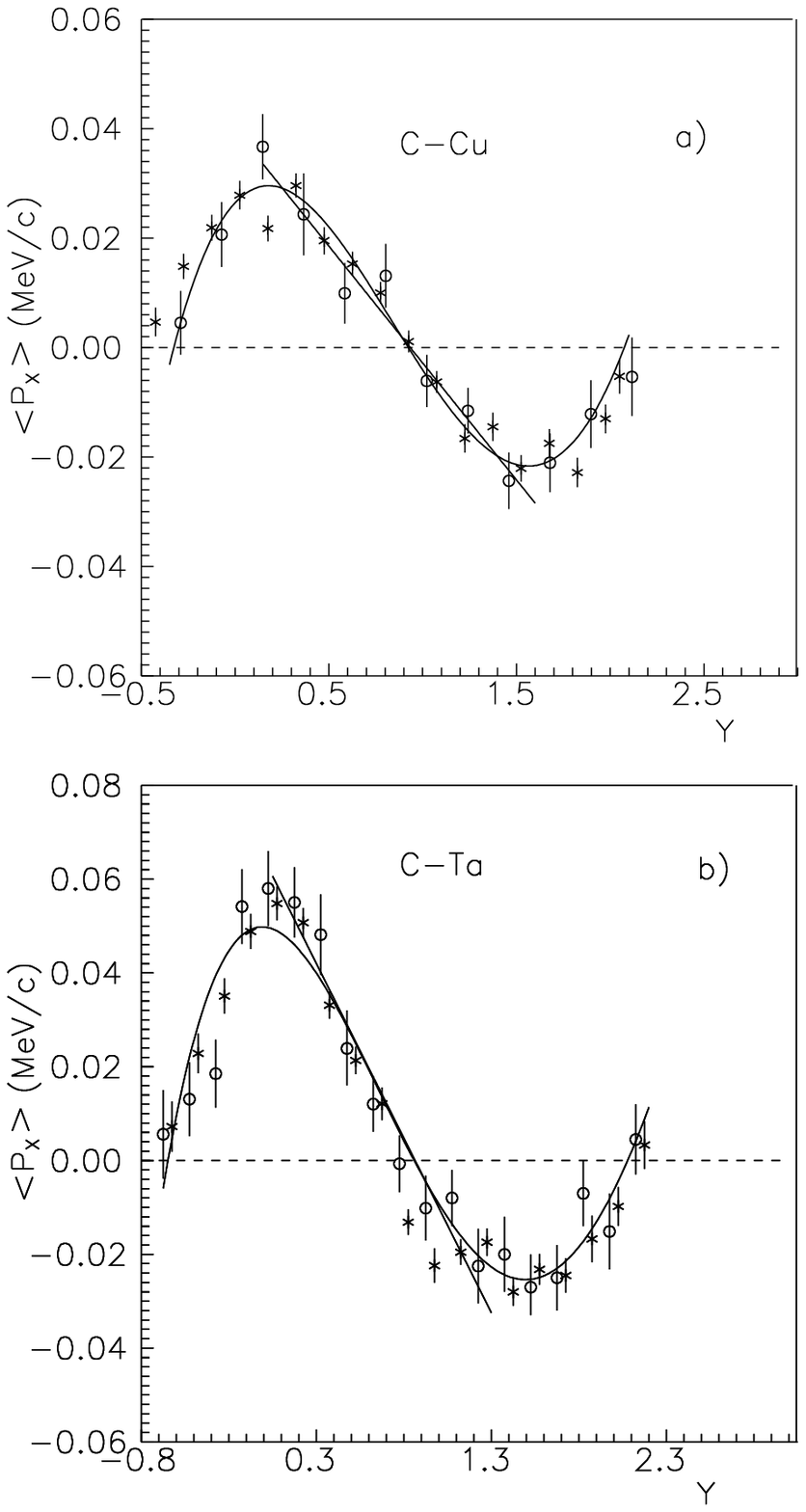}
\caption
{Average component of $\pi^-$-transverse momentum in the reaction plane, as a function of rapidity
in the lab.\ system of C-Cu (a) and C-Ta (b) collisions.  The data, corrected
for $\langle \cos{\Phi} \rangle$ from Table \ref{tab:info}, are represented by circles.
Results from the QGSM model are represented by crosses.  Straight-line stretches represent
the slope of data at midrapidity cross-over, obtained by fitting the data with a third-order polynomial
within the rapidity region of $0.15 < y < 1.60$ for C-Cu and $0.0 < y < 1.25$ for
C-Ta.  The curved lines guide the eye over data.}
\label{fig:fig8}
\end{figure}

For our systems, we have investigated the directed flow of negative pions.
The average component of pion transverse momentum in the reaction plane, evaluated using Eqs.~(\ref{eq:pxp}) and~(\ref{eq:cosfi}),
is shown in Figs.~\ref{fig:fig7} and \ref{fig:fig8}, for He-C, C-C, C-Ne, C-Cu and C-Ta collisions, respectively.  The values of flow
parameter~$F$ at cross-over, from the fits to data illustrated in Figs.~\ref{fig:fig7} and \ref{fig:fig8}, are given in Table \ref{tab:info} and
they are further plotted against the geometric mean of colliding masses in Fig \ref{fig:fig3}.
As may observed by examining the Table and by comparing Figs.~\ref{fig:fig7} and~\ref{fig:fig8} with Figs.~\ref{fig:fig1} and~\ref{fig:fig2},
the situation with the directed flow of pions is different than with the flow of protons.
The maximal transverse momentum and the flow parameter are smaller for pions than for protons.  Further,
while the average in-plane transverse momentum of protons
gradually increases in the C region as the mass of the counterpart nucleus increases from He to Ta, for pions the momentum first decreases in
magnitude and then changes sign.  Also, the flow parameter changes sign.

The reduced magnitude of average pion momentum component compared to protons has been seen before
at Bevalac \cite{gutb1,dan3,keane}, Saturne \cite{gos1}, GSI-SIS \cite{bril1} and CERN-SPS~\cite{ogi,aggar}.
From the thus-far investigated systems in the literature, the maximal in-plane momentum and flow parameter for pions appear to be the largest in magnitude
relative to protons for our C-Ta system, where $F^\pi = - 74 \pm 7 \, \text{MeV/c} $.
The direction of pion flow opposite to proton flow, termed antiflow, has been seen before in either asymmetric \cite{gos1} or symmetric
\cite{bar2,aggar,kint} systems.
However, we are unaware of an observation of the pion antiflow in strongly asymmetric systems, such as our C-Cu and C-Ta, where the pion in-plane
momentum would be simultaneously changing sign as a function of rapidity.  The cross-over in in-plane momentum
for pions is pushed towards C or nucleon-nucleon midrapidity
compared to protons
in our strongly asymmetric C-Cu and C-Ta systems.

If only the collective motion played a role in giving rise to the finite average in-plane particle momenta, the pattern of $\langle p^x \rangle$
vs rapidity should be similar for protons and pions, with $\langle p^x \rangle/m$ being of comparable magnitude for the particles \cite{bass}.
This is actually what is observed for our lighter systems, He-C, C-C and C-Ne.  The comparable values of particle $\langle p^x \rangle/m$ imply, though,
small values of pion $\langle p^x \rangle$ due to collective effects,
that can be then easily modified by an shadowing effects~\cite{bass,li1,li2}.  The latter effects can become
pronounced in more peripheral collisions of heavier systems.  Successful restricting of relative centrality in a collision relying on multiplicity gets difficult
in highly asymmetric systems.  With this, the conditions become ripe for shadowing in our C-Cu and C-Ta systems.  Quantitatively, the shadowing can
produce in-plane transverse momentum components comparable to the momenta itself and, thus, much larger than components due to
collective motion for pions~\cite{dan95}.  As can be seen from the QGSM results overlaid over the data in Figs.~\ref{fig:fig7} and \ref{fig:fig8}, the model
reproduces the pion flow-antiflow effects found in the experiment.

Given the past and present results for asymmetric systems, it could be useful to study highly asymmetric systems at higher statistics allowing for finer separations
of centrality, to illuminate when antiflow in the region of the lighter nucleus might change into flow, at antiflow persisting in the rapidity region of
heavier nucleus.

\section{ Conclusions }

Characteristics of transverse collective flow for protons and $\pi^-$ mesons emitted
from He-C, C-C, C-Ne, C-Cu and C-Ta reactions at momentum $4.2 \div 4.5$\,A\,GeV/c
(energy $3.4 \div 3.7$\,A\,GeV) have been determined.  The He-C system is the lightest in which either the directed or elliptic flow has been
ever detected.

In determining the directed flow of protons and pions,
the transverse momentum technique of Danielewicz and Odyniec has been employed.
Proton in-plane momentum appears always to maximize in the region of the lighter of the nuclei.
In the C rapidity-region,
maximal value of proton average in-plane momentum increases dramatically as
mass of the counterpart nucleus increases.  Also, the flow parameter at rapidity cross-over
increases, from  $F^p = 95 \pm 8$~MeV/c for He-C to $F^p = 178 \pm 20$~MeV/c for C-Ta.

Negative pions exhibit directed flow consistent with that for protons in He-C, C-C and C-Ne collisions.
In~those collisions, the relation between the momenta is consistent with that expected for pure effects of collective motion, i.e.\
$\langle p^x \rangle^\pi \gtrsim (m_\pi/m_p) \, \langle p^x \rangle^p$ and $F^\pi \gtrsim (m_\pi/m_p) \, F^p$.
On the other hand, for the heavier systems, C-Cu and C-Ta, pion flow turns into antiflow, with pion average in-plane momenta
becoming opposite to those for protons both in the C and counterpart nucleus rapidity regions.  The maximal pion momenta
achieve larger values than in the lighter systems, likely due to shadowing, with magnitude of pion flow parameter reaching 40\% of
that for protons, at $F^\pi = - 74 \pm 7 \, \text{MeV/c} $ for C-Ta.

In assessing the proton elliptic flow, we have combined Fourier decomposition of proton azimuthal distributions with a transverse
tensor method.  In all investigated systems at our energies, the elliptic flow points out of the reaction plane, as is characteristic
for reactions at lower beam energies.  The strength of the elliptic flow increases as mass of the system increases, from $a_2 = -0.039 \pm
0.008$ for He-C to $a_2 = -0.093 \pm 0.009$ for C-Ta.

The flow measurements have been compared to the QGSM model of collisions.  The model adequately describes flow in the measured systems.

\begin{acknowledgements}
We express our deep gratitude
to \framebox{N.~Amaglobeli} for his continuous support.  We are very grateful to
\framebox{N.~Amelin} for providing us with the QGSM code program COLLI. We~are further thankful
to V.~Uzhinski for many valuable remarks and suggestions.  One of us
(L.~Ch.) would like to thank the board of Directors of the
Laboratory of High Energies of JINR for the warm hospitality.  Finally, we thank
J.~Lukstins and O.~Rogachevsky for assistance during the preparation of
the manuscript.  This work was partially supported by the
National Science Foundation under Grant Nos.\ PHY-0245009 and PHY-0555893.
\end{acknowledgements}

\listoftables

\listoffigures

\begin{table}
\caption{Characteristics of the measured collision events and those
simulated within the Quark-Gluon String Model (QGSM), including event number~$N$ prior to multiplicity cut.\\[-1.5ex]
}
\label{tab:info}
\begin{tabular}{|c|c|c|c|c|c|}    \hline
&&& &     &   \\
\hspace*{0.5cm} & \hspace*{0.5cm} He-C \hspace*{0.5cm} &
\hspace*{0.5cm} C-C \hspace*{0.5cm} &
\hspace*{0.5cm} C-Ne \hspace*{0.5cm} &\hspace*{0.5cm} C-Cu \hspace*{0.7cm}&
    \hspace*{0.5cm} C-Ta\hspace*{0.5cm}    \\
&& &   &   & \\
\hline
&& &   &   & \\
  $N_\text{expt}$ &9737& 15962  &      723      &   663  & 2469    \\
&& &   &   & \\
\hline
&& &   &   & \\
  $N_\text{QGSM}$ & 25000& 50000  &      16337      &   5137  & 10000    \\
&& &   &   & \\
\hline
&& &   &   & \\
 $\langle \cos{ \Phi}\rangle$  & 0.866& 0.893& 0.848  &  0.784 & 0.720  \\
&& &   &   & \\
\hline
&& &   &   & \\
  $\langle \cos{ 2 \Phi}\rangle$  & 0.563 & 0.582& 0.531 &  0.559 & 0.520  \\
&& &   &   & \\
\hline
&   &  &  &&   \\
$F_\text{expt}^p$  (MeV/c) &95  $\pm$ 8 &115  $\pm$ 11 &
123  $\pm$ 12 &143 $\pm$ 15&178 $\pm$ 20\\
($k$-corrected)
&   &  &  &&   \\[1.5ex]
\hline
&& &   &   & \\
 \hspace*{.5em}  $F_\text{QGSM}^p$  (MeV/c) \hspace*{.5em} &93  $\pm$ 6 &111  $\pm$ 8 &
125  $\pm$ 8 &138 $\pm$ 11 &183 $\pm$ 12\\
&   &  &  &&   \\
\hline
&& &   &   & \\
 $F_\text{expt}^{\pi^-}$  (MeV/c) &17  $\pm$ 3 &19 $\pm$ 3 &
24  $\pm$ 5 &-43 $\pm$ 6 &-74 $\pm$ 7\\
($k$-corrected)
&   &  &  &&   \\[1.5ex]
\hline
&& &   &   & \\
  $F_\text{QGSM}^{\pi^{-}}$  (MeV/c)&18  $\pm$ 3 &18  $\pm$ 3 &
23  $\pm$ 3 &-43 $\pm$ 4 &-72 $\pm$ 4\\
&   &  &  &&   \\
\hline
\end{tabular}
\end{table}

\begin{table}
\caption{Characteristics of proton elliptic flow.}
\label{tab:eli}
\begin{tabular}{|c|c|c|c|c|}    \hline
&    &  & & \\
System & Applied Cut   & $a_{2}^\text{expt}$ &
  $R_\text{expt}$ & $a_{2}^\text{QGSM}$ \\
&  &    & & \\
\hline
     &  $-1.3 \leq y_{cm}\leq 1.3$&
 -0.039$\pm$0.008&  1.081$\pm$0.025 & -0.038$\pm$0.006\\
\cline{2-5}
&      $-1.3 \leq y_{cm} \leq 1.3$;&
  -0.049$\pm$0.010 & 1.103$\pm$0.040 &-0.047$\pm$0.007\\
He-C     &   $p^\perp \geq 0.2$ GeV/c  &   && \\
\cline{2-5}
  &  $-0.4 \leq y_{cm} \leq 2.1$&-0.062$\pm$0.011 &
1.132$\pm$0.024 &-0.056$\pm$0.007 \\
&  $p^\perp \geq 0.3$ GeV/c & & &\\
\hline
     &  $-1.3 \leq y_{cm}\leq 1.3$&
 -0.048$\pm$0.005&  1.101$\pm$0.025 & -0.048$\pm$0.004\\
\cline{2-5}
&      $-1.3 \leq y_{cm} \leq 1.3$;&
  -0.064$\pm$0.006 & 1.137$\pm$0.040 &-0.064$\pm$0.004\\
C-C     &   $p^\perp \geq 0.2$ GeV/c  &   && \\
\cline{2-5}
  &  $-0.4 \leq y_{cm} \leq 2.1$&-0.076$\pm$0.007 &
1.164$\pm$0.032 &-0.075$\pm$0.004 \\
&  $p^\perp \geq 0.3$ GeV/c & & &\\
\hline
     &  $-0.4 \leq y_{lab}\leq 2.1$&
 -0.052$\pm$0.016& 1.110$\pm$0.040 & -0.055$\pm$0.004\\
\cline{2-5}
&      $-0.4 \leq y_{lab} \leq 2.1$;&
  -0.067$\pm$0.020 & 1.144$\pm$0.039 &-0.069$\pm$0.004\\
C-Ne     &   $p^\perp \geq 0.2$ GeV/c  &   && \\
\cline{2-5}
  &  $-0.4 \leq y_{lab} \leq 2.1$&-0.080$\pm$0.020 &
1.174$\pm$0.049 &-0.081$\pm$0.004 \\
&  $p^\perp \geq 0.3$ GeV/c & & &\\
\hline
        &$-0.4 \leq y_{lab} \leq 2.1$&
 -0.072$\pm$0.013&   1.155$\pm$0.038 & -0.070$\pm$0.005\\
\cline{2-5}
&      $-0.4 \leq y_{lab} \leq 2.1$;&
      -0.077$\pm$0.015 &  1.167$\pm$0.047 &-0.080$\pm$0.005 \\
 C-Cu    &   $p^\perp\geq 0.2$ GeV/c  & &  & \\
\cline{2-5}
  &  $-0.4 \leq y_{lab} \leq 2.1$&-0.093$\pm$0.016
& 1.206$\pm$0.060 & -0.094$\pm$0.005\\
&  $p^\perp\geq 0.3$ GeV/c & & &\\
\hline
     &   $-0.3 \leq y_{lab} \leq 2.0$&
 -0.093$\pm$0.009&  1.205$\pm$0.062 &-0.095$\pm$0.004   \\
\cline{2-5}
&      $-0.3 \leq y_{lab} \leq 2.0$;&
      -0.123$\pm$0.010 & 1.280$\pm$0.065 &-0.136$\pm$0.004   \\
 C-Ta    &   $p^\perp\geq 0.2$ GeV/c  &  && \\
\cline{2-5}
  &  $-0.3 \leq y_{lab} \leq 2.0$&-0.152$\pm$0.012
& 1.358$\pm$0.075 &-0.162$\pm$0.004 \\
&  $p^{\perp}\geq 0.3$ GeV/c & & &\\
\hline
\end{tabular}
\end{table}


\end{document}